\documentclass[a4paper,12pt]{article}

\usepackage{bbold}
\usepackage{float}
\usepackage{cancel}

\usepackage{amsmath, amssymb, setspace,cite}
\usepackage{graphicx}
\usepackage{hyperref}
\usepackage{color}
\usepackage{soul}
\usepackage{multirow}
\usepackage{longtable}
\usepackage{verbatim}
\usepackage[table]{xcolor}

\graphicspath{ {./images/} } 
\vspace{0.6cm}
\date{\today} 

\usepackage{tabularx}
\usepackage{longtable} 
\usepackage{multicol}
\usepackage{multirow}
\usepackage{slashbox}
\usepackage{bigdelim}

\renewcommand{\arraystretch}{1.2}

\definecolor{light-gray}{gray}{0.90}

\begin{document}

\def\thefootnote{\fnsymbol{footnote}}

\begin{center}
\Large\bf\boldmath
\vspace*{1.cm} 
Determination of the structure of the $K \to \pi\pi\pi$ amplitudes from recent data
\unboldmath
\end{center}
\vspace{0.6cm}

\begin{center}
G.~D’Ambrosio$^{1,}$\footnote{Electronic address: gdambros@na.infn.it}, 
M.~Knecht$^{2}$\footnote{Electronic address: Marc.Knecht@cpt.univ-mrs.fr }, 
S.~Neshatpour$^{1,}$\footnote{Electronic address: neshatpour@na.infn.it}\\
\vspace{0.6cm}
{\sl $^1$INFN-Sezione di Napoli, Complesso Universitario di Monte S. Angelo,\\ Via Cintia Edificio 6, 80126 Napoli, Italy}\\[0.4cm]
{\sl $^2$Centre de Physique Th\'eorique, CNRS/Aix-Marseille Univ./Univ. de Toulon (UMR 7332)\\CNRS-Luminy Case 907, 13288 Marseille Cedex 9, France}\\

\end{center}

\renewcommand{\thefootnote}{\arabic{footnote}}
\setcounter{footnote}{0}

\vspace{1.cm}
\begin{abstract}
This Letter provides a new determination of the ten real coefficients that describe the structure of the $K\to \pi\pi\pi$ amplitudes in the limit where isospin is conserved and complex phases, due to either CP violation or final-state rescattering, are neglected. This determination is obtained through a fit to the data on the Dalitz-plot structures and partial-decay rates collected during the last twenty years by several high-precision experiments. The fitting procedure and the way the experimental data have been handled in the fit are discussed in detail. Our fit leads to a more precise determination of the coefficients describing the linear and quadratic slopes of the $K\to \pi\pi\pi$ amplitudes.
\end{abstract}

\clearpage
\section{Introduction}
The amplitudes of the $K\to\pi\pi\pi$ weak non-leptonic decay modes are usually described in terms 
of a Taylor expansion up to second order around the centre of their respective Dalitz plot 
\cite{Weinberg:1960zza,Zemach:1963bc,Devlin:1978ye,Kambor:1991ah}, see eq. (\ref{eq:ISAmps}) below.
In the limit where complex phases due to CP-violating effects or to final-state rescattering 
of the pions are neglected, 
the coefficients of the corresponding quadratic polynomials in the squares of the di-pion invariant masses 
are real.\footnote{Small imaginary parts
produced by contributions at higher orders in the chiral expansion (see the next paragraph), like the two-loop sunset-type diagram consisting
of a lowest-order $K\pi\pi\pi$ vertex connected to a six-pion vertex,
are also neglected.} If in addition isospin is assumed to be conserved, the number of coefficients is restricted to ten.
The knowledge of the numerical values of these amplitude coefficients gained through more and more 
precise experimental studies of the Dalitz-plot structures over the years has proven valuable in several respects.

\indent

First, the $K\to\pi\pi\pi$ amplitudes have been studied up to one loop precision \cite{Kambor:1991ah,Bijnens:2002vr} in the chiral expansion 
\cite{Weinberg:1978kz,Gasser:1984gg}. At this order, the coefficients of the quadratic polynomials receive contributions from low-energy constants of the 
next-to-leading order effective Lagrangian in the weak sector \cite{Kambor:1989tz,Esposito-Farese:1990inm,Ecker:1992de}.
Estimates of these low-energy constants \cite{Ecker:1990in,Isidori:1991ya,Ecker:1992de,DAmbrosio:1997ctq} through resonance saturation are much more model-dependent than the
corresponding estimates \cite{Ecker:1989yg,Ecker:1988te} of the one-loop low-energy constants in the 
strong sector. The information provided by the experimental studies of the Dalitz plots is thus a manner to 
determine at least some combinations of these low-energy constants 
\cite{Kambor:1991ah,Bijnens:2002vr} and to constrain the arbitrary parameters that
enter their phenomenological estimates within resonance-saturation approaches. In a somewhat similar vein, two-loop 
representations of the $K\to\pi\pi\pi$ amplitudes have recently been constructed \cite{Kampf:2019bkf} in the isospin limit 
(and in some cases also with isospin-breaking effects due to the difference between charged and neutral pion masses),
combining the chiral counting with general properties like analyticity, unitarity and crossing. These two-loop 
representations involve a polynomial part, representing the contributions from the low-energy constants,
and whose coefficients can for instance be fixed through the knowledge of the amplitude coefficients.

\indent  

Second, the amplitude coefficients also appear as parameters involved in phenomenological descriptions of amplitudes for
radiative decays of a kaon into a pion. For this class of decays (see e.g. section VI.D of the review \cite{Cirigliano:2011ny} for a 
comprehensive discussion and an extensive bibliography), the amplitude vanishes at lowest-order in the chiral expansion \cite{Ecker:1987fm,Ecker:1987qi,Ecker:1987hd}, and the first non-trivial 
contribution occurs at one loop. It often happens, though, that due to the chiral power counting, the one-loop approximation does not allow 
to generate all the invariant amplitudes allowed by Lorentz invariance and electromagnetic gauge invariance. One-loop predictions 
for the decay rate and/or the decay distribution are then often at variance with experimental measurements, requiring the theoretical 
description to go beyond this approximation and to include corrections to it. In order to circumvent a full-fledged two-loop 
calculation, these corrections are most of the time restricted to the inclusion of unitarity corrections. 
These in turn involve a $K\to\pi\pi\pi$ vertex beyond its lowest-order expression, which is then taken as given by its 
polynomial expansion to second order.
Illustrative examples where a procedure of this kind has been implemented include the decay modes of the following list:
$K_L\to\pi^0\gamma\gamma$ \cite{Cappiello:1992kk,Cohen:1993ta,Kambor:1993tv}, $K^\pm\to\pi^\pm\gamma\gamma$   \cite{DAmbrosio:1996cak},
$K\to\pi\ell^+\ell^-$  \cite{DAmbrosio:1998gur}, $K^\pm\to\pi^\pm\gamma e^+ e^-$   \cite{Gabbiani:1998tj}, and
$K_L\to\pi^0\gamma \ell^+\ell^-$ \cite{Donoghue:1997rr,Donoghue:1998ur}. Experimental data are then fitted to these 
beyond-one-loop theoretical expressions with the $K\to\pi\pi\pi$ amplitude coefficients taken as external parameters.

\indent  

Finally, one might even envision the possibility to abandon the low-energy expansion altogether, and to start from a dispersive 
representation of the amplitudes, as recently discussed in refs. \cite{Kambor:1993tv,Colangelo:2016ruc} for the $K_S\to \gamma \gamma^{(*)}$ amplitudes,
or in ref. \cite{DAmbrosio:2018ytt} for the case of the $K\to\pi\ell^+\ell^-$ decay modes. In the second case,
this would require to reconstruct the relevant $K\pi\to\pi\pi$ partial-wave amplitudes using a unitarization procedure,
like, for instance, the dispersive method devised quite some time ago by N.~N.~Khuri and S.~B.~Treiman \cite{Khuri:1960zz}.
More complete and more elaborate implementations of this dispersive set-up have been described in detail in the 
recent literature \cite{Kambor:1995yc,Anisovich:1996tx,Colangelo:2018jxw,Descotes-Genon:2014tla,Guo:2015zqa,Gasser:2018qtg}.
Here, the $K\to\pi\pi\pi$ amplitude coefficients can be used in order to fix the subtraction constants that are required for making the dispersive representations sufficiently convergent.
It certainly remains to be seen whether such a much more ambitious program can be carried out and lead 
to better theoretical descriptions and understanding of the radiative $K\to\pi$ transitions, but a good knowledge of the  
$K\to\pi\pi\pi$ amplitude coefficients will definitely be one of its essential inputs.

\indent 

The last compilation to date of the $K\to\pi\pi\pi$ amplitude coefficients as extracted from experimental data 
on the decay distributions and branching fractions goes back to ref. \cite{Bijnens:2002vr}. In the meantime, the experimental 
situation has witnessed tremendous improvements in precision, with several new measurements of both the
energy dependence of the $K\to\pi\pi\pi$ Dalitz plots 
\cite{Ajinenko:2002mg,Akopdzhanov:2005nb,NA482:2007exu,KTeV:2008gel}
and of the $K\to\pi\pi\pi$ partial widths
\cite{KLOE:2003new,KTeV:2004hpx,NA48:2005uiw,KLOE:2005vdt,KLOEKLOE-2:2014tsu},
as shown by the corresponding entries in the recent issue of the Review of Particle Physics (PDG) 
\cite{ParticleDataGroup:2020ssz}.\footnote{Actually, we have been using
the updated on-line version available at https://pdg.lbl.gov/  and that will become available in print as
R.L. Workman et al. (Particle Data Group), to be published in Prog. Theor. Exp. Phys. 2022, 083C01 (2022).
For our needs,  the two versions present identical entries.}
It is thus about time to provide an updated version of table 5 of ref. \cite{Bijnens:2002vr} that includes
the experimental progress made since its publication twenty years ago. This is the purpose of this Letter. Its remaining content is organized as follows:
section \ref{sec:framework} provides the essential theoretical tools needed for this study: the expressions of all 
$K\to\pi\pi\pi$ amplitudes as second-order polynomials, and the link with the experimentally observed Dalitz-plot
structures and decay widths.
Section \ref{sec:procedure} describes the fit procedure we have followed. Our results are presented and discussed in section \ref{sec:results}.
Two additional tables, serving the purpose of illustrating some observations 
or comments made in the main text, have been gathered in a short appendix.

\section{Theoretical framework}\label{sec:framework}

Under the conditions stated at the beginning of the preceding section,
the amplitudes for the $K\to \pi\pi\pi$ amplitudes are parameterized
in terms of ten real coefficients $\alpha_{1,3}$, $\beta_{1,3}$, $\gamma_3$, $\zeta_{1,3}$, $\xi_{1,3}$, $\xi'_3$ as~\cite{Kambor:1991ah,Bijnens:2002vr}
(the subscripts  ``1'' or ``3'' refer to terms induced through $\Delta I = 1/2$ or $\Delta I = 3/2$ transitions, respectively)

\begin{subequations}\label{eq:ISAmps}
\begin{align}
A(K_L\to  \pi^0\pi^0\pi^0)&= -3({\alpha_1}+{\alpha_3})
         -3 ({\zeta_1}-2{\zeta_3})\left(Y^2+\frac{1}{3}X^2\right)\,,
\label{eq:ISAmps:a}\\[6pt]
A(K_L\to  \pi^+\pi^-\pi^0)&= ({\alpha_1}+{\alpha_3})-({\beta_1}+{\beta_3})Y
         + ({\zeta_1}-2{\zeta_3})\left(Y^2+\frac{1}{3}X^2\right)\nonumber\\[-6pt]&
         +({\xi_1}-2{\xi_3})\left(Y^2-\frac{1}{3}X^2\right) \,,
\label{eq:ISAmps:b}\\[6pt]
A(K^\pm\to  \pi^0\pi^0\pi^\pm)&= -\left({\alpha_1}-\frac{1}{2}{\alpha_3}\right)
           +\left({\beta_1}-\frac{1}{2}{\beta_3}-\sqrt{3}{\gamma_3}\right)Y
         - ({\zeta_1}+{\zeta_3})\left(Y^2+\frac{1}{3}X^2\right)
\nonumber\\&
         -({\xi_1}+{\xi_3}+{\xi_3^\prime})\left(Y^2-\frac{1}{3}X^2\right) \,,
\label{eq:ISAmps:c}\\[6pt]
A(K^\pm\to  \pi^\pm\pi^\pm\pi^\mp) &= 2\left({\alpha_1}-\frac{1}{2}{\alpha_3}\right)
           +\left({\beta_1}-\frac{1}{2}{\beta_3}+\sqrt{3}{\gamma_3}\right)Y
         +2 ({\zeta_1}+{\zeta_3})\left(Y^2+\frac{1}{3}X^2\right)
\nonumber\\[-4pt]&
         -({\xi_1}+{\xi_3}-{\xi_3^\prime})\left(Y^2-\frac{1}{3}X^2\right) \,,
\label{eq:ISAmps:d}\\[6pt]
A(K_S\to  \pi^+\pi^-\pi^0)&= \frac{2}{3}\sqrt{3}\,{\gamma_3} X- \frac{4}{3}{\xi_3^\prime} XY\,, \label{eq:ISAmps:e}
\end{align}
\end{subequations}
with $X\equiv (s_2-s_1)/(m_{\pi^+}^2)$ and $Y\equiv (s_3-s_0)/(m_{\pi^+}^2)$ and  the Lorentz invariant kinematic parameters $s_i$ are given by
\begin{align}
s_i = (k-p_i)^2\,, 
&& 
 s_0 = \frac{1}{3}(m_K^2 + m_{\pi^1}^2 + m_{\pi^2}^2 + m_{\pi^3}^2)\,
\end{align}
where $k^\mu$ and $p_i^\mu$ correspond to the momenta of the kaon and of the pions, respectively, with $p_i^2=m_{\pi^i}^2$. 
Although the above representation of the amplitudes rests on invariance under isospin transformations, 
the corresponding physical masses for the kaon and pions are used for the numerical evaluation of $s_0$ and of the phase space integrals.
However, by convention the charged pion mass is used in the definitions of the variables $X$ and $Y$ in all cases. In addition, since
CP-violating effects are not taken into account, we do not distinguish between the masses of $K_L$ and $K_S$ and
use the $K^0$ mass for the neutral kaons.

\indent  

For any of these decay modes, the doubly-differential decay distribution
$d^2 \Gamma(X,Y)/dXdY$ with respect to the variables $X$ and $Y$ is directly proportional to the
modulus squared of the corresponding amplitude. Experimentally, the latter is expressed as 
(obviously, the present discussion excludes the decay $K_S\to  \pi^+\pi^-\pi^0$)
\begin{align}
\vert A(X,Y) \vert^2  = \vert A(0,0) \vert^2 [ 1 + g Y + h Y^2 + k X^2]  .
\end{align}
A fit to the experimental Dalitz-plot distribution then provides values for the linear and quadratic slopes $g$, $h$ and $k$. In order to fix the absolute normalization $\vert A(0,0) \vert^2$, one also needs a measurement of the corresponding decay width. It is useful to remember that the values of the amplitudes (\ref{eq:ISAmps}) at the centre of the Dalitz plot ($X=Y=0$) only involve the coefficients $\alpha_1$ and $\alpha_3$, so that the latter
are expected to be the most sensitive to the values of the decay widths.

There are overall 15 relevant physical observables: 4 decay rates corresponding to
$K_L\to  \pi^0\pi^0\pi^0$, $K_L\to  \pi^+\pi^-\pi^0$, 
$K^\pm \to \pi^0\pi^0\pi^\pm$ and $K^\pm\to  \pi^\pm\pi^\pm\pi^\mp$, 10 slope parameters
$g$, $h$ and $k$ corresponding to these decays 
(for $K_L\to  \pi^0\pi^0\pi^0$, Bose symmetry requires $g = 0$ and $k = h/3$) and finally we have Re$\left(\lambda\right)$ 
corresponding to the $K_S \to \pi^+ \pi^- \pi^0$ decay. 
Here $\lambda$ is the complex parameter describing the interference of the
$l = 1$ component of the $K_S \to \pi^+ \pi^- \pi^0$ decay amplitude with the 
$l = 0$ component of the $K_L \to \pi^+ \pi^- \pi^0$ amplitude 
($X_{\rm lim}(Y)$, $Y_{\rm min}$ and $Y_{\rm max}$ denote 
the limits of the phase space)
%
\begin{align}\label{eq:lambda}
\lambda&= \frac{\displaystyle\int_{Y_{\rm min}}^{Y_{\rm max}}dY\int_0^{X_{\rm lim}(Y)}dX
\;A_{L}^{*(l=0)}(X,Y) A_S^{(l=1)}(X,Y)}{\displaystyle\int_{Y_{\rm min}}^{Y_{\rm max}}dY\int_0^{X_{\rm lim}(Y)}dX
\;|A_{L}(X,Y)|^2}\,.
\end{align} 

\indent

In order to extract values of the ten amplitude coefficients in (\ref{eq:ISAmps}) (collectively denoted as~$\theta$) from a fit to the experimental data on $K\to \pi\pi\pi$ decays we use the  method of maximum likelihood within a frequentist approach based on a Gaussian approximation for the likelihood function ${\cal L}(\theta)$:
\begin{align}\label{eq:chi2}
    -2\, \ln {\cal L}(\theta) = \chi^2(\theta) = \sum_{i,j=1}^{15}\left(O_i^{\rm th}(\theta)-O_i^{\rm exp}\right)\,C^{-1}_{i,j}\,\left(O_j^{\rm th}(\theta)-O_j^{\rm exp}\right)
\end{align}
where $C_{i,j}$ is the $(i,j)$ element of the experimental covariance matrix and $O_i^{\rm th}(\theta)$ and $O_i^{\rm exp}$ correspond to the 
theoretical prediction and experimental measurement of the $i$-th observable, respectively. These theoretical predictions are obtained
upon using the formulas (\ref{eq:ISAmps}) of the amplitudes. In the expressions for the moduli squared $\vert A(X,Y) \vert^2$,
or in the product of amplitudes appearing in eq. (\ref{eq:lambda}), only the terms at most quadratic in $X$ and $Y$ are kept.
As a test, we performed the fit using the data available at the time of publication of ref. \cite{Bijnens:2002vr}. The result reproduces to an excellent precision the values found in table 5 of ref. \cite{Bijnens:2002vr} and shown in the fourth column of table~\ref{tab:ScaledFit_OurAvg}, including the value of the $\chi^2$. For the sake of comparison, we also show, in the second and third columns of table~\ref{tab:ScaledFit_OurAvg}, the values quoted in the two earlier references \cite{Devlin:1978ye} and \cite{Kambor:1991ah}, respectively.

\indent 

Proceeding in the same way, but
using now as experimental input the entries of the latest issue of the PDG, results in a much larger value of the $\chi^2$ and a 
much downgraded quality of the fit.
Details are reported in appendix~\ref{app:PDGavg_PDGfit}. Understanding the origin and the reason for such an unexpected bad outcome 
necessitates to have a closer look at the
way the data are handled and combined in the PDG. This investigation, which is described in the next section, 
will eventually also provide a path toward a result with a much more 
acceptable value of the $\chi^2$. The reader only interested in our final result will find it in the last column of table~\ref{tab:ScaledFit_OurAvg}.

\section{Fitting procedure and data}\label{sec:procedure}

Besides Re$\left(\lambda\right)$, for which we use the average from refs.~\cite{Zou:1996ks,CPLEAR:1997snn,NA48:2005uiw,CPLEAR:1998nkj}, the values for
the other observables listed in the first column of table~\ref{tab:ScaledInputs_OurAvg} are provided by the PDG's weighted averages of the most up-to-date
and/or statistically most significant experimental measurements. 
These values are shown in the second column of table~\ref{tab:ScaledInputs_OurAvg}. 
Sometimes, only a single experiment is retained by the PDG in these ``averages''. When several experiments are used, no correlations are provided by the PDG.
For some of these average values, the PDG has applied a scale factor on the resulting uncertainties. For reasons that will become clear shortly,
we have removed these scale factors in the uncertainties shown in the second column of table~\ref{tab:ScaledInputs_OurAvg}.
In the case of the partial widths the PDG also reports the results from a constrained fit  
where information from other kaon decay modes are taken into account while enforcing the sum of the corresponding branching ratios to 
add up to unity. In this case, the correlations among the decay widths (separately for the decay modes of charged or neutral kaon) are also provided. In the 
sequel, the set of experimental observables where the averages or fitted values of the partial widths are being used will be 
referred to as the \textit{PDG average} (PDG-avg) or the \textit{PDG fit} (PDG-fit), respectively.

\indent

The fit of the amplitude coefficients when taking the PDG-avg as the experimental input results in a large 
$\chi^2/{\rm dof} = 17.4/5$ 
(details are reported in Appendix~\ref{app:PDGavg_PDGfit}).
The situation is even worse when employing the PDG-fit as experimental input, resulting in $\chi^2/{\rm dof} = 44.9/5$, an increase
which is not unexpected as the PDG-fit values for the partial widths have smaller uncertainties as compared to PDG-avg.  
The source of these large $\chi^2$ values can be better understood by checking the share of each observable in the total $\chi^2$. 
In table~\ref{tab:PDG_fit_avg_remove_paper} of appendix~\ref{app:EachObsChi2} we give the individual contributions to the total $\chi^2$ where we have 
used the fitted external parameters ($\bar{\theta}$) to calculate $O_i^{\rm th}(\bar{\theta})$ for each summand in eq.~\ref{eq:chi2}. 
From this table it is clear that the large $\chi^2$ is mostly due to the decay widths.
However, it is not obvious whether the problem is due specifically to these observables or merely reflects an overall tension within the data.

\indent  

Looking for possible explanations, we first notice that the value of $A(0,0)$ involves two different combinations,
$2\alpha_1 - \alpha_3$ for the charged kaon and $\alpha_1 + \alpha_3$ for the neutral one. This suggests that an origin for the large $\chi^2$ values 
could perhaps be found in a tension between the decay widths of the charged kaon on the one hand and of the neutral kaon on the other hand.
However, this turns out not to be the case, as doing separate fits to  the 
amplitude coefficients (with an alternative set 
of combinations as defined in Ref.~\cite{DAmbrosio:1994vba}) for the charged and neutral kaon decays does not improve the fit. Moreover, the sum of 
the two separate $\chi^2$ values gives back the value of the $\chi^2$ obtained previously for the fit to all data simultaneously. 
Interestingly enough, the main source of tension rather seems to lie between the values of two partial widths given for each type of kaon, neutral or charged.
As shown in table~\ref{tab:PDG_fit_avg_remove_paper}, removing either of the two decay rates (whether for the neutral or the charged kaon), the large $\chi^2$ 
from the other decay rate (for the same kaon) also 
becomes completely relaxed, thus improving the fit substantially and suggesting that the origin of the problem could be due to tensions between the different experimental measurements of the branching fractions and/or the total widths. 
For a naive check of the impact of the decay widths on the fit,
we have redone the fit by doubling the uncertainties of the decay width  while keeping the errors of all the other observables unchanged. 
As shown in table~\ref{tab:PDG_avg_fit_ExternalFit}, the $\chi^2$ is more than halved by this procedure.
As to the amplitude coefficients themselves, the only substantial difference between the two outputs shown in table~\ref{tab:PDG_avg_fit_ExternalFit} 
lies in the doubling of the uncertainties on $\alpha_1$ and $\alpha_3$, which was to be expected,
since these two quantities are the most sensitive to the decay widths. 

\indent

A further indication that the explanation for the large $\chi^2$ may find its origin in some  tensions between different data is provided by the fact that 
values quoted by the PDG for the corresponding branching ratios and lifetimes already have quite large scale factors.  Among the 15 observables we 
consider for the fit, the following relevant observables contributing to the set PDG-avg have a scale factor (SF) larger than unity:
\begin{itemize}
 \item branching ratio of $K_L\to \pi^0\pi^0\pi^0$: SF $=2.0$
 \item weighted average for the total lifetime of $K^+$: SF $=1.9$
 \item branching ratio of $K^+\to \pi^0\pi^0\pi^+$: SF $=1.2$
 \item linear slope $g$ of $K_L\to \pi^+\pi^-\pi^0$: SF $=1.5$
 \item quadratic slope $k$ of $K^+\to \pi^0\pi^0\pi^+$: SF $=2.5$
\end{itemize}
It is worth noting that $\Gamma(K_L\to \pi^0\pi^0\pi^0)$, which has the most sizable contribution to the large $\chi^2$ of the fit, 
is dependent on BR($K_L\to \pi^0\pi^0\pi^0$) and the value of the latter is affected by a $\sim 2\sigma$
tension between the KTeV~\cite{KTeV:2004hpx} and KLOE~\cite{KLOE:2005vdt} measurements.

\indent 

Assuming that the large $\chi^2/{\rm dof}$ is due to tensions between the various experimental 
results\footnote{From a strictly logical perspective the necessity to introduce scale factors could, at least partly, also be due to a possible shortcoming in the description of the amplitudes by simple polynomials with only ten independent real coefficients, as given by the expressions (\ref{eq:ISAmps}). We will address the possible effects of e.g. $\pi\pi$ rescattering phases or violations of isospin symmetry in section \ref{sec:results}.\label{fnote:assumptions}} that enter the fit, 
we consider a scaled fit for the amplitude coefficients.
In general, to deal with inconsistent data in a least-square fit a scale factor can be introduced for the experimental data~\cite{ParticleDataGroup:2020ssz}.
The justification behind introducing a scale factor is that in principle one or more of the experimental data have underestimated errors and when it is not possible to determine which data set is the source of the discrepancy, the scale factor addresses this ignorance. In our case,  the discrepancy is observed in the fit of several parameters (as opposed to the weighted average of one physical parameter). To address this issue we employ the method described in the introduction of ref. \cite{ParticleDataGroup:2020ssz} for performing a scaled~fit.

\begin{table}[t]
\renewcommand{\arraystretch}{1.3}
\begin{center}
\scalebox{0.85}{
\begin{tabular}{|c||c|c|c|c|}
\hline
\multirow{ 2}{*}{observable} & {\footnotesize $O_i^{\rm exp} \pm \delta O_i^{\rm exp}$}  & {\footnotesize $\bar{O}_i \pm\delta \bar{O}_i$} & \multirow{ 2}{*}{SF} & {\footnotesize $O_i^{\rm exp} \pm \delta O_i^{\prime\,{\rm exp}}$}  \\[-4pt]
& (orig. err.) &  (predicted result) & & (scaled err.) \\
\hline\hline
$\Gamma(K_L \to \pi^0 \pi^0 \pi^0)\cdot 10^{18}$ 	& $ 2.5417 \pm 0.0198 $			& $ 2.5994 \pm 0.0129 $	& 3.8	& $ 2.5417 \pm 0.0759 $	\\
$\Gamma(K_L \to \pi^+ \pi^- \pi^0)\cdot 10^{18}$ 	& $ 1.6200 \pm 0.0102 $			& $ 1.5950 \pm 0.0079 $	& 3.8	& $ 1.6200 \pm 0.0392 $	\\
$\Gamma(K^+ \to \pi^0 \pi^0 \pi^+)\cdot 10^{18}$ 	& $ 0.9438 \pm 0.0128 $			& $ 0.9115 \pm 0.0058 $	& 2.8	& $ 0.9438 \pm 0.0363 $	\\
$\Gamma(K^+ \to \pi^+ \pi^+ \pi^-)\cdot 10^{18}$ 	& $ 2.9590 \pm 0.0213 $			& $ 2.9864 \pm 0.0190 $	& 2.8	& $ 2.9590 \pm 0.0604 $	\\ \hline\hline
$h(K_L \to \pi^0 \pi^0 \pi^0)$ 	& $ 0.0006 \pm 0.0012 $			& $ 0.0000 \pm 0.0011 $	& 1.8	& $ 0.0006 \pm 0.0021 $	\\ \hline
$g(K_L \to \pi^+ \pi^- \pi^0)$ 	& $ 0.678 \pm 0.005 $			& $ 0.679 \pm 0.008 $	& 1.0	& $ 0.678 \pm 0.005 $	\\
$h(K_L \to \pi^+ \pi^- \pi^0)$ 	& $ 0.076 \pm 0.006 $			& $ 0.082 \pm 0.006 $	& 2.6	& $ 0.076 \pm 0.015 $	\\
$k(K_L \to \pi^+ \pi^- \pi^0)$ 	& $ 0.0099 \pm 0.0015 $			& $ 0.0110 \pm 0.0012 $	& 1.3	& $ 0.0099 \pm 0.0019 $	\\ \hline
$g(K^\pm \to \pi^0 \pi^0 \pi^\pm)$ 	& $ 0.626 \pm 0.007 $			& $ 0.622 \pm 0.021 $	& 1.0	& $ 0.626 \pm 0.007 $	\\
$h(K^\pm \to \pi^0 \pi^0 \pi^\pm)$ 	& $ 0.052 \pm 0.008 $			& $ 0.069 \pm 0.082 $	& 1.0	& $ 0.052 \pm 0.008 $	\\
$k(K^\pm \to \pi^0 \pi^0 \pi^\pm)$ 	& $ 0.0054 \pm 0.0014 $			& $ 0.0070 \pm 0.0013 $	& 2.7	& $ 0.0054 \pm 0.0037 $	\\ \hline
$g(K^\pm\to  \pi^+\pi^-\pi^\pm)$ 	& $ -0.21134 \pm 0.00017 $			& $ -0.21134 \pm 0.00691 $	& 1.0	& $ -0.21134 \pm 0.00017 $	\\
$h(K^\pm\to  \pi^+\pi^-\pi^\pm)$ 	& $ 0.0185 \pm 0.0004 $			& $ 0.0185 \pm 0.0091 $	& 1.0	& $ 0.0185 \pm 0.0004 $	\\
$k(K^\pm\to  \pi^+\pi^-\pi^\pm)$ 	& $ -0.00463 \pm 0.00014 $			& $ -0.00464 \pm 0.00021 $	& 1.0	& $ -0.00463 \pm 0.00014 $	\\ \hline
Re$\left[\lambda (K_S \to \pi^+ \pi^- \pi^0)\right]$ 	& $ 0.0334 \pm 0.0052 $			& $ 0.0340 \pm 0.0012 $	& 1.0	& $ 0.0334 \pm 0.0052 $	\\ \hline
\end{tabular}
}
\end{center}
\caption{\small 
The relevant data used in the scaled fit.
In the second column, the PDG-avg set without scale factors is given. The third column gives the theoretical prediction of each observable 
$\bar{O}_i \equiv  O_i^{\rm th}(\bar{\theta})$ employing the fitted parameters $\bar{\theta}$ obtained from the fit to the data in the second column. 
The scale factors are given in the fourth column, using eq.~\ref{eq:Si} and the two left columns. The last column which gives the data used in the 
scaled fit contains the same data as the second column but the uncertainties have been multiplied by the corresponding scale factor from the fourth column.
\label{tab:ScaledInputs_OurAvg}}
\end{table}

\indent

In this procedure, in order to calculate the scale factor for each experimental input, as a first step the fit should be done with all relevant input data endowed with their 
``original errors''\footnote{It does not make much sense to add scale factors on uncertainties already endowed with a scale factor, therefore we 
start with the PDG data set \emph{without} the scale factors that are already applied there, and thus determine our own scale factors.} -- see second column in 
table~\ref{tab:ScaledInputs_OurAvg} where we have taken all measurements 
entering the PDG-avg (except for Re($\lambda$), which is not provided 
by the PDG compilation) without the scale factors. 
Once this ``unscaled'' fit is done, we calculate the scale factor $S_i$ for each of the experimental values as
\begin{align}\label{eq:Si}
    S_i^2  = \frac{\big(O_i^{\rm exp} - \bar{O}_i \big)^2}{\big(\delta O_i^{\rm exp}\big)^2 - \big(\delta \bar{O}_i\big)^2}   ,
\end{align}
where $O_i^{\rm exp}$ and $\delta O_i^{\rm exp}$ stand for the experimental central value and the corresponding (unscaled) uncertainty, respectively.
$\bar{O}_i \equiv  O_i^{\rm th}(\bar{\theta})$ refers to the observable $O_i$ as predicted by the fit, and $\delta \bar{O}_i$ its uncertainty taking into account correlations 
among the fitted parameters $\bar{\theta}$ (third column in table~\ref{tab:ScaledInputs_OurAvg}). 
The scale factor is then considered as the larger of $S_i$ and unity for 
each observable (fourth column of table~\ref{tab:ScaledInputs_OurAvg}).
As expected the largest scale factor is obtained for the two decay widths of the neutral kaon.%
\footnote{Interestingly enough, the value SF $=3.8$ of the scale factor 
for $\Gamma(K_L\to\pi^0\pi^0\pi^0)$ we find in table~\ref{tab:ScaledInputs_OurAvg} is rather close to what one would find in the case where 
the constraint from $\sum{\rm BR}_i=1$ is not enforced in the PDG fit (R. Bonventre, C.-J. Lin, P. Zyla, private communication).}

\indent 

In the final step we redo the fit with the scaled uncertainties obtained by multiplying each uncertainty by the corresponding scale 
factor ($\delta O_i^{\prime{\rm exp}}$ as given in the last column of table~\ref{tab:ScaledInputs_OurAvg}).
To report the fitted parameters of this ``scaled fit'' we follow PDG's policy of not having the scale factors impact the central values, 
thus we give the central values of the fitted parameters from the original (unscaled) fit.

\section{Results, discussion, conclusion}\label{sec:results}

Our main result consists of the outcome of our scaled fit, given in the last column of table~\ref{tab:ScaledFit_OurAvg}, and of 
the correlation matrix for the corresponding uncertainties that we provide in table~\ref{tab:PDGavg_CorrMatrix}
for completeness. This scaled fit produces a good value for the corresponding $\chi^2$.
To appreciate the effect of scaled data on the fit, we also give the resulting scale factor for the uncertainty of the fitted parameters,
obtained by dividing the uncertainties of the fitted parameters from the scaled fit by those from the unscaled fit.

\begin{table}[ht]
\renewcommand{\arraystretch}{1.3}
\begin{center}
\rowcolors{3}{}{light-gray}
\scalebox{0.9}{
\begin{tabular}{|c|c|c|c||cc|}
\hline
{amplitude} & Devlin et \emph{al.}  & {Kambor et \emph{al.}} & {Bijnens et \emph{al.} }  & \multirow{ 2}{*}{Our scaled fit} & \multirow{ 2}{*}{SF}\\[-6pt]
coefficient  & (Ref.~\cite{Devlin:1978ye}) &  (Ref.~\cite{Kambor:1991ah}) & (Ref.~\cite{Bijnens:2002vr})  &  &  \\
\hline\hline
$\alpha_1$	&	$91.4\pm0.24    $	&	$91.71\pm0.32   $	&	$93.16\pm0.36   $	&	$ 92.80 \pm 0.64 $	& 2.9	\\
$\alpha_3$     	&	$-7.14\pm0.36   $	&	$-7.36\pm0.47   $	&	$-6.72\pm0.46   $	&	$ -7.45 \pm 0.79 $	& 3.2	\\
$\beta_1$	&	$-25.83\pm0.41  $	&	$-25.68\pm0.27  $	&	$-27.06\pm0.43  $	&	$ -26.46 \pm 0.22 $	& 1.6	\\
$\beta_3$        	&	$-2.48\pm0.48   $	&	$-2.43\pm0.41   $	&	$-2.22\pm0.47   $	&	$ -2.50 \pm 0.29 $	& 1.6	\\
$\gamma_3$	&	$2.51\pm0.36    $	&	$2.26\pm0.23    $	&	$2.95\pm0.32    $	&	$ 2.78 \pm 0.10 $	& 1.0	\\
$\zeta_1$	&	$-0.37\pm0.11   $	&	$-0.47\pm0.15   $	&	$-0.40\pm0.19   $	&	$ -0.11 \pm 0.03 $	& 1.7	\\
$\zeta_3$       	&	 ---             	&	$-0.21\pm0.08   $	&	$-0.09\pm0.10   $	&	$ -0.05 \pm 0.03 $	& 1.8	\\
$\xi_1$   	&	$-1.25\pm0.12   $	&	$-1.51\pm0.30   $	&	$-1.83\pm0.30   $	&	$ -1.20 \pm 0.13 $	& 1.7	\\
$\xi_3$  	&	 ---             	&	$-0.12\pm0.17   $	&	$-0.17\pm0.16   $	&	$ 0.10 \pm 0.10 $	& 1.6	\\
$\xi_3^\prime$	&	 ---             	&	$-0.21\pm0.51   $	&	$-0.56\pm0.42   $	&	$ -0.07 \pm 0.16 $	& 1.8	\\\hline\hline
$\chi^2/{\rm dof}$        	&	$ 12.8/3          $	&	$10.3/2       $	&	$5.4/5 $	&	\multicolumn{2}{c|}{$ 5.18/5$ \;($30.66/5$)}	 	\\\hline 
\end{tabular}
}
\end{center}
\caption{\small
The fit result of the amplitude coefficients (in units of $10^{-8}$). Following PDG's procedure for our scaled fit, only the uncertainties are taken from the scaled fit while the central values are taken from the fit to the data with the original errors. The given uncertainties already include the scale factor, the correlations among uncertainties are given table~\ref{tab:PDGavg_CorrMatrix}. In the last column of the last row, the number in the parenthesis corresponds to the value of the $\chi^2/{\rm dof}$ before the data have been scaled.
\label{tab:ScaledFit_OurAvg}}
\end{table}

\indent 

As compared to the previous determination of the amplitude 
coefficients, the central values we obtain are compatible with those of ref. \cite{Bijnens:2002vr} within the quoted uncertainties.
However, despite sometimes significant values of the scale factors, the uncertainties on the coefficients corresponding to the linear and quadratic slopes are reduced substantially, reflecting the improvement in the quality of the experimental data on the Dalitz-plot structures.
Nevertheless, the contributions $\zeta_3$, $\xi_3$, $\xi_3^\prime$ from the $\Delta I = 3/2$ transitions 
to the quadratic slopes, besides being reduced with respect to the corresponding $\Delta I =1/2$ contributions, remain with large relative uncertainties, both features being merely reflections of the $\Delta I =1/2$ rule in the $K\to\pi\pi\pi$ sector.
Finally, in the case of $\alpha_1$ and $\alpha_3$ we obtain larger uncertainties than in the previous study \cite{Bijnens:2002vr}. This results from
the large scale factor that the fit produces for these two coefficients, pointing back to
the tension between the decay-width measurements.

\begin{table}[htb]
\renewcommand{\arraystretch}{1.3}
\begin{center}
\scalebox{0.75}{
\begin{tabular}{|c|rrrrrrrrr|}
\hline
\multicolumn{10}{|c|}{Correlation matrix of the scaled fit } \\\hline
& $\alpha_3$ & $\beta_1$ & $\beta_3$ & $\gamma_3$ & $\zeta_1$ & $\zeta_3$ & $\xi_1$ & $\xi_3$ & $\xi_3^\prime$\\\hline
$\alpha_1$	 & $-37.83$	 & $-79.87$	 & $21.04$	 & $22.71$	 & $-7.35$	 & $2.72$	 & $-7.25$	 & $-0.67$	 & $-1.25$ \\
$\alpha_3$	 &  	 & $22.21$	 & $-78.46$	 & $-16.91$	 & $-4.20$	 & $8.30$	 & $0.44$	 & $7.28$	 & $0.55$ \\
$\beta_1$	 &  	 &  	 & $-12.74$	 & $-65.40$	 & $7.56$	 & $-3.93$	 & $14.93$	 & $4.20$	 & $11.06$ \\
$\beta_3$	 &  	 &  	 &  	 & $47.40$	 & $5.45$	 & $-8.55$	 & $-4.26$	 & $-11.92$	 & $-7.81$ \\
$\gamma_3$	 &  	 &  	 &  	 &  	 & $-1.15$	 & $-0.36$	 & $-17.07$	 & $-11.68$	 & $-19.87$ \\
$\zeta_1$	 &  	 &  	 &  	 &  	 &  	 & $-89.31$	 & $16.01$	 & $-22.14$	 & $-0.42$ \\
$\zeta_3$	 &  	 &  	 &  	 &  	 &  	 &  	 & $-16.61$	 & $22.94$	 & $-0.29$ \\
$\xi_1$	 &  	 &  	 &  	 &  	 &  	 &  	 &  	 & $-5.70$	 & $78.63$ \\
$\xi_3$	 &  	 &  	 &  	 &  	 &  	 &  	 &  	 &  	 & $54.46$ \\
\hline
\end{tabular}
}
\end{center}
\caption{Correlations among the fitted parameters for our scaled fit (in percentage).
\label{tab:PDGavg_CorrMatrix}}
\end{table}

\indent  

As mentioned in the introduction, several combinations of the amplitude coefficients appear as external parameters in
phenomenological parameterizations of the amplitudes for some rare kaon decay modes. The values and the uncertainties on these
specific combinations can be obtained from the contents of table~\ref{tab:ScaledFit_OurAvg} and table~\ref{tab:PDGavg_CorrMatrix}.
For the convenience of the interested reader, we provide, in table \ref{tab:combinations_OurScaledFit},
some of these values and indicate the processes where they are relevant
(the corresponding references are given in the introduction).

\begin{table}[htb]
\renewcommand{\arraystretch}{1.3}
\begin{center}
\scalebox{0.9}{
\begin{tabular}{|c|r|c|}
\hline
\multicolumn{1}{|c|}{combination} & \multicolumn{1}{c|}{value ($10^{-8}$)}& decay mode \\\hline
$2\alpha_1 - \alpha_3$		& $ 193.05 \pm 1.74 $				&\rdelim\}{4}{5.5cm}[$K^\pm \to \pi^\pm \gamma \gamma, K^\pm \to \pi^\pm\gamma \ell^+ \ell^-$]	\\
$2\zeta_1 - \xi_1$		& $ 0.99 \pm 0.14 $					& \\
$4\zeta_1 + \xi_1$		& $ -1.63 \pm 0.20 $					& \\
$\beta_1 -\tfrac{1}{2}\beta_3+\sqrt{3}\gamma_3$		& $ -20.40 \pm 0.18 $	&\quad\rdelim\}{2}{5.5cm}[$K^\pm \to \pi^\pm \ell^+ \ell^-$]\\
$2(\xi_1 + \xi_3 -\xi_3^\prime)$		& $ -2.05 \pm 0.06 $					&\\
$\alpha_1 + \alpha_3$ 		& $ 85.35 \pm 0.81 $					&\rdelim\}{4}{5.5cm}[$K_L\to \pi^0 \gamma \gamma, K_L\to \pi^0 \gamma \ell^+ \ell^-$]\\
$\beta_1 + \beta_3$ 		& $ -28.96 \pm 0.34 $					& \\
$\zeta_1 -2\zeta_3 + \xi_1 - 2\xi_3$ 		& $ -1.41 \pm 0.28 $					& \\
$\zeta_1 -2\zeta_3 - \xi_1 + 2\xi_3$ 		& $ 1.41 \pm 0.23 $					& \\
\hline 
\end{tabular}
}
\end{center}
\caption{\small
Specific combinations of interest as used in different decay modes such as $K^+ \to \pi^\pm \gamma \gamma$, $K^\pm \to \pi^\pm \gamma \ell^+ \ell^-$, $K^+ \to \pi^+ \ell\ell$, $K_L\to \pi^0 \gamma \gamma$ and $K_L\to \pi^0 \gamma \ell^+ \ell^-$  calculated from the 10-dimensional scaled fit. 
The uncertainties include the correlations among the parameters. 
\label{tab:combinations_OurScaledFit}}
\end{table}

\indent 

The expressions (\ref{eq:ISAmps}) of the 
$K\to\pi\pi\pi$ amplitudes hold only under a certain set of conditions that were listed
at the beginning of the introduction. We know that
in real life these conditions are only approximately satisfied. As already briefly mentioned above, see footnote~\ref{fnote:assumptions},
these approximations could also be responsible, at least partly, for the scale factors that were necessary in order
to produce a satisfactory result for the fit of the amplitude coefficients.
This issue thus certainly deserves to be briefly discussed.
The first condition that needs to be met is the absence of CP violation.
Now, CP-violating effects in $K\to\pi\pi\pi$ decays have been studied both from a theoretical \cite{DAmbrosio:1991oli} and 
from an experimental \cite{NA482:2007ucr} point of view,
and were found to be tiny, well below the level where they could have a visible impact on the description of the amplitudes, given the present experimental uncertainties on the decay rates and Dalitz-plot parameters. 
As to the second condition,
namely that $\pi\pi$ rescattering phases can be neglected, it can also be tested quantitatively.
Indeed, the imaginary parts generated by $\pi\pi$ rescattering at one loop
have been computed in ref. \cite{DAmbrosio:1994vba}.\footnote{They can also
be obtained from the full one-loop calculation of ref. \cite{Bijnens:2002vr},
but the way they were presented in ref. \cite{DAmbrosio:1994vba} is more convenient
for our purpose. Moreover the authors of the former reference checked that
numerically they found agreement between their results and those from the latter.}.
Computing the corresponding shifts in the Dalitz-plot variables and redoing
the fit with the corrected parameters $g$, $h$ and $k$, we find no change
in the scale factors that need to be applied, and the shifts in the central
values of the resulting amplitude coefficients are quite small and 
more than generously covered by the uncertainties.
Finally, isospin-breaking effects have been studied in detail 
at the one-loop level in the low-energy expansion in a 
series of articles \cite{Bijnens:2004ku,Bijnens:2004vz,Bijnens:2004ai}.
In order to assess their possible effect on the quality of the fit,
we have redone the fit, using either the PDG-fit or the PDG-avg data
(in the last case including the original scale factors), adding  the isospin-breaking 
corrections as given in tables 3 and 4 of ref. \cite{Bijnens:2004vz} to the observables. 
This leads only to marginal changes in the quality of the fit, 
with the $\chi^2$ decreasing (increasing) by about 15\% for PDG-fit (PDG-avg) 
as compared to the value given in the left part of table \ref{tab:PDG_avg_fit_ExternalFit}. 
At the level of the amplitude coefficients, the changes are well within the uncertainties
shown in this same table, except for $\alpha_1$ and $\alpha_3$,
whose variations are somewhat larger than one standard deviation. This 
is in line with the observation made by the authors of ref.
\cite{Bijnens:2004vz} that the main effect of the isospin-breaking 
contributions lies in the values of the squares of the decay amplitudes 
at the centre of the Dalitz plot, with much milder incidences
on the slopes. It should be stressed that the entries of tables 3 
and 4 of ref. \cite{Bijnens:2004vz} are given without uncertainties,
and that they were obtained upon putting some
combinations of unknown low-energy constants to zero at some reference
scale. Their effect could become numerically significant if they 
were given some typically expected values. It seems thus difficult to
reliably give a  more quantitative assessment of how much isospin-breaking
corrections would affect the scale factors that we have determined in table
\ref{tab:ScaledFit_OurAvg}.

\indent 

To conclude, from what is known about their size, $\pi\pi$ rescattering or isospin-breaking effects by 
themselves do not explain the scale factors applied in the averages
of several experimental inputs to the fit, and listed in section \ref{sec:procedure}. 
These scale factors are mainly driven by some tension between different experimental
determinations of the decay widths. But part of 
the larger scale factors we encounter in our scaled fit could actually also reflect the fact that
the parameterizations (\ref{eq:ISAmps}) start to have difficulties in correctly representing the data 
as they become more precise. This is actually already the case for the latest data on
the $K^+\to \pi^0\pi^0\pi^+$ decay mode published by the NA48/2 collaboration \cite{NA482:2005wht}, where a clear
isospin-breaking effect, in the form of a cusp in the distribution with respect to the invariant 
mass squared of the two neutral pions, has been observed and has even been used,
with some theoretical input \cite{Cabibbo:2004gq,Cabibbo:2005ez,Colangelo:2006va,Gasser:2011ju,Bissegger:2008ff},
in order to improve the experimental determination of the pion-pion scattering lengths in the S wave.
Such a feature in the Dalitz-plot distribution goes clearly beyond the simple polynomial 
parameterization of eq. (\ref{eq:ISAmps}, and the corresponding data can not be directly included
in our fit. A similar cusp has also been seen in the Dalitz-plot distribution of the decay mode
$K_L\to \pi^0\pi^0\pi^0$ by the KTeV experiment \cite{KTeV:2008gel} but it is much less pronounced, 
so that we could still include these data in our fit without any particular problem.

\indent  

\section*{Acknowledgements}
We are grateful to several colleagues who, through insightful discussions and correspondence,
have been helpful in improving our understanding of some important issues  concerning the data and their
treatment: E. Goudzovski, M. Koval, and A. Shaikhiev from the NA62 
collaboration on the one hand, and, on the other hand, 
R. Bonventre, C.-J. Lin, and P. Zyla from the Particle Data Group.
We would also like to thank F. Ambrosino, A. Bizetti, C. Lazzeroni, M. Moulson, and G. Ruggiero for discussions and for their interest in this work.
The work of G. D. and of S. N. was supported in part by the INFN research initiative Exploring New Physics (ENP).
The work of M. K. has received partial support from the Excellence Initiative of Aix-Marseille
University~--~A$^*$MIDEX, a French ``Investissement d'Avenir" program (AMX-19-IET-008 - IPhU).

\indent

\indent

\clearpage
\begin{appendix}

\section{Inputs from PDG average or fit}\label{app:PDGavg_PDGfit}
In this appendix we give the analysis for the fit to $K\to 3\pi$ data when using PDG average or the PDG fit with the uncertainties as quoted in PDG (including the scale factors).
\subsection{External parameters fit}\label{app:origFit}

In table \ref{tab:PDG_avg_fit_ExternalFit} we show the results of the fit for the amplitude coefficients
for the two data sets PDG-fit and PDG-avg, first when keeping the original uncertainties, and next when the 
uncertainties on the partial widths have been doubled while keeping the uncertainties on the remaining data unchanged.
This doubling substantially improves the value of the $\chi^2$, but at the expense of the accuracy in the determination
of the amplitude coefficients $\alpha_1$ and $\alpha_3$ that give the values of the amplitudes at the centre of the Dalitz-plot.

\begin{table}[htb]
\renewcommand{\arraystretch}{1.3}
\begin{center}
\rowcolors{4}{}{light-gray}
\scalebox{0.85}{
\begin{tabular}{|c|c|c|}
\multicolumn{3}{c}{} \\\hline
 & PDG-fit & PDG-avg \\
\hline\hline
$\alpha_1$	&	$92.63 \pm 0.16$	&	$92.87 \pm 0.23$	\\	
$\alpha_3$     	&	$-7.46 \pm 0.21$	&	$-7.13 \pm 0.27$	\\	
$\beta_1$	&	$-26.53 \pm 0.15$	&	$-26.50 \pm 0.16$	\\	
$\beta_3$        	&	$-2.63 \pm 0.24$	&	$-2.53 \pm 0.24$	\\	
$\gamma_3$	&	$2.80 \pm 0.09$	&	$2.80 \pm 0.09$	\\	
$\zeta_1$	&	$-0.11 \pm 0.02$	&	$-0.11 \pm 0.02$	\\	
$\zeta_3$       	&	$-0.05 \pm 0.02$	&	$-0.05 \pm 0.02$	\\	
$\xi_1$   	&	$-1.36 \pm 0.12$	&	$-1.35 \pm 0.12$	\\	
$\xi_3$  	&	$0.04 \pm 0.07$	&	$0.03 \pm 0.07$	\\	
$\xi_3^\prime$	&	$-0.30 \pm 0.16$	&	$-0.30 \pm 0.16$	\\	\hline\hline
$\chi^2/{\rm dof}$        	&	$ 44.92 /5 $	&	$17.38/5$	\\	
\hline
\end{tabular}
}
\quad
\rowcolors{4}{}{light-gray}
\scalebox{0.85}{
\begin{tabular}{|c|c|c|}
\multicolumn{3}{c}{} \\\hline
 &  PDG-fit & PDG-avg.\\
 \hline\hline
$\alpha_1$ 	&	 $ 92.65 \pm  0.31$ 	&	 $92.87 \pm 0.46$ \\ 
$\alpha_3$      	&	 $ -7.44 \pm  0.42$ 	&	 $-7.12 \pm 0.54$ \\ 
$\beta_1$ 	&	 $ -26.43 \pm  0.17$ 	&	 $-26.47 \pm 0.19$ \\ 
$\beta_3$         	&	 $ -2.43 \pm  0.26$ 	&	 $-2.48 \pm 0.28$ \\ 
$\gamma_3$ 	&	 $ 2.80 \pm  0.09$ 	&	 $2.80 \pm 0.10$ \\ 
$\zeta_1$ 	&	 $ -0.11 \pm  0.02$ 	&	 $-0.10 \pm 0.02$ \\ 
$\zeta_3$        	&	 $ -0.05 \pm  0.02$ 	&	 $-0.06 \pm 0.02$ \\ 
$\xi_1$    	&	 $ -1.35 \pm  0.12$ 	&	 $-1.35 \pm 0.12$ \\ 
$\xi_3$   	&	 $ 0.03 \pm  0.07$ 	&	 $0.02 \pm 0.07$ \\ 
$\xi_3^\prime$ 	&	 $ -0.30 \pm  0.16$ 	&	 $-0.30 \pm 0.16$ \\ \hline\hline
$\chi^2/{\rm dof}$         	&	 $  15.05/5 $ 	&	 $7.81/5$ \\ 
\hline
\end{tabular}
}
\end{center}
\caption{Comparison of the fit results for the $K\to\pi\pi\pi$ amplitude coefficients when the inputs are taken from 
the two data sets PDG-fit and PDG-avg. In the left part the uncertainties on the partial decay widths have been kept as 
given by the PDG, in the right part these uncertainties have been doubled.
\label{tab:PDG_avg_fit_ExternalFit}}
\end{table}

\clearpage
\subsection{\texorpdfstring{Contribution of each observable to the total {\boldmath$\chi^2$}}{Contribution of each observable to the total chi-squared}}
\label{app:EachObsChi2}
In table~\ref{tab:PDG_fit_avg_remove_paper} below we give the contribution 
of each observable to the $\chi^2$ using the full data or when removing only 
one of the decay widths from the fit.
\begin{table}[h]
\renewcommand{\arraystretch}{1.3}
\begin{center}
\scalebox{0.55}{
\hspace{-0.75cm}
\begin{tabular}{|c||cc||c|c|c|c||cc||c|c|c|c|}
\hline
& \multicolumn{6}{c||}{Experimental input for decay widths: PDG fit} & \multicolumn{6}{c|}{Experimental input for decay widths: PDG average}  \\[1pt]\hline
\multirow{ 2}{*}{\backslashbox{Observable}{Removing}} &  & None & $\Gamma(K_L^{000})$ & $\Gamma(K_L^{+-0})$  & $\Gamma(K^{00+})$ & $\Gamma(K^{++-})$ &  & None & $\Gamma(K_L^{000})$ & $\Gamma(K_L^{+-0})$  & $\Gamma(K^{00+})$ & $\Gamma(K^{++-})$\\[-2pt]
 & input & $\chi^2$ & $\chi^2$ & $\chi^2$  & $\chi^2$ & $\chi^2$ & input & $\chi^2$ & $\chi^2$ & $\chi^2$  & $\chi^2$ & $\chi^2$ \\
\hline\hline
$\Gamma(K_L \to \pi^0 \pi^0 \pi^0)\cdot 10^{18}$ 	& $2.5112 \pm 0.0201$&17.69	& –	& 0.00	& 17.68	& 17.69	& $2.5417 \pm 0.0352$ & $5.36$	& –	& $0.00$	& $5.36$	& $5.37$	\\[-4pt]
\small ${\scriptstyle \updownarrow}$ correlation ${\scriptstyle \updownarrow}$	& 
\footnotesize 42.0\% &
\footnotesize 8.92	& –	& –	& 
\footnotesize  8.93	&  
\footnotesize 8.92	& & 	&    	&    	&    	&    	\\[-4pt]
$\Gamma(K_L \to \pi^+ \pi^- \pi^0)\cdot 10^{18}$ 	& $1.6133 \pm 0.0101$&6.38	& 0.00	& –	& 6.39	& 6.37	& $1.6200 \pm 0.0102$ & $1.20$	& $0.00$	&  –	& $1.20$	& $1.18$	\\ \hline
$\Gamma(K^+ \to \pi^0 \pi^0 \pi^+)\cdot 10^{18}$ 	& $0.9356 \pm 0.0121$&4.95	& 4.95	& 4.95	& –	& 0.00	& $0.9438 \pm 0.0150$ & $5.12$	& $5.13$	& $5.13$	& –	& $0.00$	\\[-4pt]
\small ${\scriptstyle \updownarrow}$ correlation ${\scriptstyle \updownarrow}$	& 
\footnotesize 3.5\% &
\footnotesize 0.10	& 
\footnotesize  0.10	&  
\footnotesize 0.10	& –	& –	& & 	&    	&    	&    	&    	\\[-4pt]
$\Gamma(K^+ \to \pi^+ \pi^+ \pi^-)\cdot 10^{18}$ 	& $2.9686 \pm 0.0127$&0.41	& 0.41	& 0.40	& 0.00	& –	& $2.9590 \pm 0.0218$ & $1.00$	& $1.00$	& $1.00$	& $0.00$	&  –	\\ \hline\hline
$h(K_L \to \pi^0 \pi^0 \pi^0)$ 	&  $0.0006\pm0.0012 $ &0.54	& 0.13	& 0.13	& 0.54	& 0.54	& $0.0006 \pm 0.0012$ & $0.20$	& $0.13$	& $0.13$	& $0.20$	& $0.20$	\\ \hline
$g(K_L \to \pi^+ \pi^- \pi^0)$ 	&  $0.678\pm0.008 $ &0.74	& 0.16	& 0.16	& 0.74	& 0.74	& $0.678 \pm 0.008$ & $0.01$	& $0.16$	& $0.16$	& $0.01$	& $0.01$	\\
$h(K_L \to \pi^+ \pi^- \pi^0)$ 	&  $0.076\pm 0.006 $ &1.47	& 0.81	& 0.81	& 1.47	& 1.48	& $0.076 \pm 0.006$ & $0.95$	& $0.81$	& $0.81$	& $0.95$	& $0.95$	\\
$k(K_L \to \pi^+ \pi^- \pi^0)$ 	&  $0.0099\pm 0.0015 $ &0.68	& 0.46	& 0.46	& 0.68	& 0.69	& $0.0099 \pm 0.0015$ & $0.51$	& $0.46$	& $0.46$	& $0.51$	& $0.50$	\\ \hline
$g(K^\pm \to \pi^0 \pi^0 \pi^\pm)$ 	&  $0.626\pm0.007 $ &0.07	& 0.07	& 0.09	& 0.10	& 0.12	& $0.626 \pm 0.007$ & $0.08$	& $0.07$	& $0.09$	& $0.10$	& $0.16$	\\
$h(K^\pm \to \pi^0 \pi^0 \pi^\pm)$ 	&  $0.052\pm0.008 $ &1.07	& 1.07	& 1.07	& 1.07	& 1.07	& $0.052 \pm 0.008$ & $1.07$	& $1.07$	& $1.07$	& $1.07$	& $1.15$	\\
$k(K^\pm \to \pi^0 \pi^0 \pi^\pm)$ 	&  $0.0054\pm0.0035 $ &1.86	& 1.86	& 1.84	& 1.83	& 1.82	& $0.0054 \pm 0.0035$ & $1.86$	& $1.86$	& $1.84$	& $1.83$	& $1.71$	\\ \hline
$g(K^\pm \to \pi^\pm \pi^+ \pi^-)$ 	&  $-0.21134\pm0.00017 $ &0.00	& 0.00	& 0.00	& 0.00	& 0.00	& $-0.21134 \pm 0.00017$ & $0.00$	& $0.00$	& $0.00$	& $0.00$	& $0.00$	\\
$h(K^\pm \to \pi^\pm \pi^+ \pi^-)$ 	&  $0.0185\pm0.0004 $ &0.00	& 0.00	& 0.00	& 0.00	& 0.00	& $0.0185 \pm 0.0004$ & $0.00$	& $0.00$	& $0.00$	& $0.00$	& $0.00$	\\
$k(K^\pm \to \pi^\pm \pi^+ \pi^-)$ 	&  $-0.00463\pm0.00014 $ &0.00	& 0.00	& 0.00	& 0.00	& 0.00	& $-0.00463 \pm 0.00014$ & $0.00$	& $0.00$	& $0.00$	& $0.00$	& $0.00$	\\ \hline
Re$\left[\lambda (K_S \to \pi^+ \pi^- \pi^0)\right]$ 	&  $0.0334\pm 0.0052 $ &0.04	& 0.03	& 0.09	& 0.03	& 0.08	& $0.0334 \pm 0.0052$ & $0.03$	& $0.02$	& $0.07$	& $0.02$	& $0.07$	\\ \hline\hline
Total $\chi^2$ 	& &44.92	& 10.05	& 10.12	& 39.46	& 39.51	&  & $17.38$	& $10.72$	& $10.77$	& $11.24$	& $11.31$	\\
\hline
\end{tabular}
}
\end{center}
\caption{\small
Experimental values and their individual contribution to the total $\chi^2$ considering the best fit value of the fitted parameters based on experimental inputs from PDG-fit or PDG-avg.
Each individual $\chi^2$ contribution is given for the fit when the the full data (indicated by ``None'') has been assumed or when removing only one of the decay widths. For the PDG-avg experimental inputs no correlations are available. For the PDG-fit where correlations are available the correlated $\chi^2$ contributions are also given (when their values are at least 0.01).
\label{tab:PDG_fit_avg_remove_paper}}
\end{table}

\end{appendix}

\clearpage

\end{document}